\documentclass[%
 reprint,
amsmath,amssymb,
aps,
prb,
]{revtex4-2}
\usepackage{amsmath}
\usepackage{physics}
\usepackage{amssymb}
\usepackage{graphicx}
\usepackage{float}
\usepackage{tabularx}
\usepackage{xcolor}
\usepackage{graphicx}
\usepackage{dcolumn}
\usepackage{bm}

\usepackage{hyperref}

\usepackage{ulem}

\newcommand{\bk}{\mathbf{k}}
\newcommand{\bq}{\mathbf{q}}

\newcommand{\vF}{v_{\!F}}
\newcommand{\ImPi}{\mathrm{Im}\,\Pi}
\newcommand{\sgn}{\mathrm{sgn}}
\newcommand{\im}{\mathrm{Im}}

\begin{document}

\preprint{APS/123-QED}

\title{Umklapp corrections to Landau damping and conditions for non-trivial modifications to quantum critical transport}
\author{Vibhu Mishra}
\email{vibhu.mishra@uni-goettingen.de}

\affiliation{%
 Institute for Theoretical Physics, Georg-August-Universität Göttingen, Friedrich-Hund-Platz 1, 37077 Göttingen,Germany
}%




\date{\today}

\begin{abstract}

We compute the particle--hole bubble for an Ising-nematic metal when the critical Fermi surface approaches the Brillouin zone boundary for $d=2$ dimensions.
We find two qualitatively distinct contributions:
i)~the standard antipodal piece, which gives $\Pi_{\rm{ATP}}(\bq, i\Omega)\propto\Omega/q$ and
ii)~an additional umklapp piece from electrons near the zone boundary, 
which gives $\Pi_{\rm{U}}(\bq, i\Omega)\propto \Omega^\alpha$ at the minimum umklapp momentum $q\approx \Delta_q$ with $\alpha = 2/3 $ or $1/2$ depending on the temperature $T$.
At high $T$ when $\alpha = 1/2$, the minimum $T$ for the activation of linear/quasi-linear in $T$ resistivity, which is expected to be $T_U \propto \Delta_q^3$ from $z=3$ criticality, could potentially get reduced to $T_U \propto \Delta_q^4$ due to the $\sqrt{\Omega}$ term and discuss why we find only one hyper-specific scenario where this possibility might be realized.
For $d=3$ the umklapp contribution gives $\Pi_{\rm{U}}\sim \Omega$ irrespective of $T$ therefore $T_U$ is not modified in this case.

\end{abstract}

\maketitle


\section{Introduction}
\label{sec:intro}

There is a peculiar aspect baked into Landau's Fermi liquid theory.
On one hand, the derivation of Landau's mass renormalization formula that connects the effective mass of the quasi-particle to Landau parameters (that characterize fermion interactions) explicitly requires Galilean invariance and in fact there is no analogous derivation for fermions on a crystal lattice \cite{coleman2015introduction, landau1957theory}. 
On the other hand, the $T^2$ resistivity that is said to be the hallmark of Fermi liquid metallicity requires umklapp back-scattering which is only possible on a lattice \cite{Hartnoll_Umklapp, Hartnoll_Lucas_Sachdev_Holographic}.
This aspect is easy to understand as a cloud of electrons in an external electric field glides across effortlessly as there is no impedance to the motion leading to a perfect conductor.
It doesn't matter how the electrons interact with one another and whether or not quasi-particles exist.
In reality, the presence of impurities and a crystal lattice circumvent this conclusion and allow for currents to decay leading to nonzero resistivity.

It has been argued that umklapp is not general enough to explain the much less well understood strange metal phenomena.
The interesting transport properties do not survive down to zero temperatures and a spatially disordered interaction was proposed as the mechanism for strange metal behavior \cite{Patel_Sachdev_Condctivity, Patel_Sachdev_Universal_Theory}.
In recent works \cite{JvD_PAM1, JvD_PAM2}, strange metal behavior was observed on a numerical simulation of the Periodic Anderson Model in $d=3$ dimensions (the microscopic model for heavy fermions) in the vicinity of (but not right at) the quantum critical point.
Since the model lives on a lattice and has no disorder whatsoever, it shows completely intrinsic (disorder-free) strange metal behavior.
It was shown in \cite{LinearT_HigherD} that the spatially disordered interaction of \cite{Patel_Sachdev_Condctivity, Patel_Sachdev_Universal_Theory} does not lead to strange metallicity for $d>2$ therefore it's not unreasonable to assume that lattice effects are at play as suggested at the end of \cite{JvD_PAM2}.

The effective model for the heavy fermion systems at criticality involves a conduction and a spinon fermi surface interacting via hybridization fluctuations \cite{Pepin_Paul_1, Paul_pepin_2, Paul_Pepin_3, Patel_Altman_Kondo}.
The model lives on a continuum therefore an  interplay of potential disorder and hybridization is required as the mechanism of $T \ln T$ resistivity which is at odds with the results from \cite{JvD_PAM1, JvD_PAM2} which is a clean model on a lattice so the effects of umklapp scattering on the clean effective model become a worthwhile question to investigate.

In two dimensions, umklapp transport due to a critical scalar field was thoroughly investigated in \cite{Wang_Berg, Lee_umklapp} who found linear/quasi-linear in $T$ resistivity above a transition temperature $T_U$. 
A relativistic bosonic mode coupled to a Fermi surface (FS) experiences Landau damping which is captured by a dressed bosonic propagator of the form \cite{Sachdev_QPT, Sachdev_Phases}
\begin{equation}
\label{eq:boson_prop}
D^{-1}(\bq,i\Omega) = q^2 + \Omega^2 + r  - \Pi(\bq,i\Omega).
\end{equation}
There are obvious overlaps and differences between this simple  Ising-nematic model and critical heavy fermions.
The differences come from the fact that the hybridization fluctuation model has a light $c$ electron and a heavy $f$ electrons with generically mismatched Fermi surfaces which give rise to additional energy-momentum scales absent from the Ising-nematic system with a single FS.
The overlap comes from the fact that the boson propagators in both cases are nearly identical at criticality which is why here we focus on the simpler model.

Within the random phase approximation, $\Pi(\bq,i\Omega)$ is built up of particle-hole bubbles summed over the entire FS.
The standard derivation of $\Pi(\bq,i\Omega)$ starts by noting that for any $\bq,$ the dominant contribution comes from the points on the FS which have tangents parallel to $\bq$.
For convex fermi surfaces there are only two such points, called the anti-podal patch contributions and the calculation gives ($q= |\bq|$)
\begin{equation}
    \Pi_{\text{ATP}}(\bq,i\Omega) \propto |\Omega|/q \sim \gamma_{\bq} |\Omega|
\end{equation}which dominates over the $\Omega^2$ term in $D(\bq,i\Omega)$ and in the $r \rightarrow 0$ limit gives rise to the $z=3$ criticality \cite{Sachdev_Phases, Sachdev_QPT}.
The derivation assumes that the FS is small and far away from the Brillouin zone (BZ) boundaries.

When the FS is large enough to approach the BZ boundary, 
umklapp scattering channels open: 
an electron near the zone face can scatter across the boundary 
to the opposite side of the FS.  
It was noted in \cite{Wang_Berg} that such processes ``modify the coefficient $\gamma_{\bq}$ of the Landau damping term\ldots\ 
This can lead to a breakdown of the $z=3$ quantum critical scaling, which relies on the relation $\gamma_{\bq}\sim 1/q$.''
In the current work, we pursue this line of reasoning further.

We perform an explicit patch calculation of the umklapp contribution to $\Pi(\bq, i\Omega) = \Pi_{\text{ATP}}(\bq,i\Omega) + \Pi_{\text{U}}(\bq,i\Omega)$ 
for $d=2$ critical Ising-nematic metal.  
We find that the umklapp piece does not simply renormalize $\gamma_{\bq}$.
Instead, it introduces a qualitatively new frequency dependence $\Pi_{\mathrm{U}} \propto \Omega^\alpha$, where $\alpha = 2/3, \, 1/2$ at low and high temperatures respectively at about the minimum umklapp momentum $q \sim\Delta_q$ and is separated from $q \sim 0$ by a frequency gap.

As a consequence $z=3$ is stable at $T \to 0 \implies \alpha = 2/3$ for any finite $\Delta_q$.
At higher $T$ for $\alpha = 1/2$, there is a possibility that the minimum temperature for the activation of critical umklapp transport which requires $q \sim \Delta_q$ is lowered from $T_U \propto \Delta_q^3$ (as suggested purely from $z=3$ criticality \cite{Wang_Berg, Lee_umklapp}) to $T_U \propto \Delta_q^4$ due to the $\sqrt{\Omega}$ piece.

We discuss necessary conditions such that the lowered activation temperature scale emerges self-consistently from the fully dressed boson propagator.
We show that despite the potentially dominating $\sqrt{\Omega}$ behavior, it is only in the so called naive large-$N$ limit \cite{Sachdev_Chowdhury_Review} (with $N$ fermion flavors and a single scalar field) studied in \cite{Wang_Berg} that we expect $\Pi_{\rm{U}}$ to make any serious non-trivial changes. 
We also show that for $d=3$, $\Pi_{\text{U}} \sim \Omega$ irrespective of $T$ therefore it never competes with $\Pi_{\text{ATP}}$ so despite the original motivation, the heavy fermion criticality remains completely unmodified by the correction.

The article is structured as follows, in Sec.[\ref{sec:setup}] we evaluate the umklapp contribution to the boson self energy.
Then in Sec.[\ref{sec:Implications}], we discuss the conditions under which $\Pi_{\mathrm{U}}$ can possibly lead to a modification of $T_U$ which is shown at the level of scattering rate in Sec.[\ref{sec:Scattering_Rate}].
Finally we end in Sec.[\ref{sec:summary}] outlining why we expect that for physically realistic systems, the effects of $\Pi_{\rm{U}}$ would be benign even in the case of $d=2$.

\section{Particle Hole Bubble}
\label{sec:setup}

We evaluate the simple p-h bubble using the free fermion propagator \cite{Sachdev_Phases}
\begin{equation}
\begin{aligned}
        \Pi(\bq, i\Omega) 
        &= -g^2\int \frac{d^2k \, d\omega}{(2 \pi)^3} \frac{1}{i(\omega + \Omega) - \varepsilon_{\bk+ \bq}} \cdot \frac{1}{i\omega - \varepsilon_{\bk}}
        \\
        &= g^2 \int \frac{d^2k}{(2 \pi)^2} \frac{n_F(\varepsilon_{\bk+\bq}) - n_F(\varepsilon_{\bk})}{i\Omega + \varepsilon_{\bk} - \varepsilon_{\bk + \bq}},
\end{aligned}
\end{equation}
where $g$ is the fermion boson coupling strength $g \int \phi c^{\dagger} c$ and $n_F(x)$ is the Fermi distribution function.

The imaginary part of the self-energy for real frequencies is obtained by $i\Omega \rightarrow \Omega + i\eta$ giving
\begin{equation}
\label{eq:ImPi}
\ImPi(\bq,\Omega) = \frac{-g^2}{4 \pi}\int\!d^2k\,
\big[n_F(\varepsilon_{\bk+\bq}) - n_F(\varepsilon_{\bk})\big]\,
\delta\!\big(\Omega + \varepsilon_{\bk} - \varepsilon_{\bk+\bq}\big)\,.
\end{equation}
Performing a Taylor expansion of $n_F(\varepsilon+\Omega)-n_F(\varepsilon)\approx \Omega\,\partial_\varepsilon n_F$ 
(valid for $\Omega\ll\varepsilon_F$ the Fermi energy) and taking $T\to 0$ 
so that $\partial_\varepsilon n_F \to -\delta(\varepsilon)$ gives
\begin{equation}
\label{eq:ImPi_T0}
\frac{\ImPi(\bq,\Omega)}{\Omega} 
= \frac{g^2}{4\pi}\int d^2k\;\delta(\varepsilon_{\bk})\;
\delta\!\big(\Omega + \varepsilon_{\bk} - \varepsilon_{\bk+\bq}\big)\,.
\end{equation}  
The first delta function restricts $\bk$ to the FS and the second enforces energy conservation.  
Generically these two constraints fix $\bk$ to isolated points, 
and the result is a sum over such solutions weighted by inverse Jacobians.

\subsection{Antipodal Patch Contribution}
\label{subsec:antipodal}

\begin{figure}
	\centering
		\includegraphics[scale=1.0]{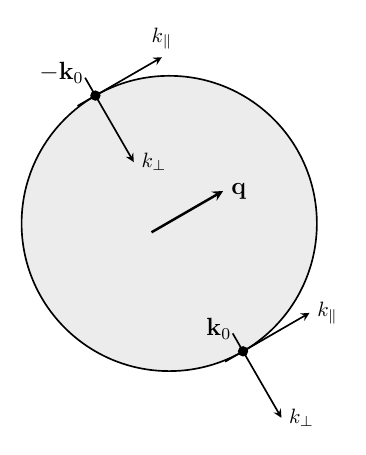}
	\caption{The two antipodal patches contributing to $\Pi_{\text{ATP}}$.}
	\label{fig:ATP}
\end{figure}

We briefly review the standard calculation to establish the mechanism that produces $1/q$.
The dominant contribution comes from the two antipodal points $\pm\bk_0$ in Fig.[\ref{fig:ATP}] where $\mathbf{v}_F\perp \bq$.
Without loss of generality we take $\bq = (0, q_{\parallel})$.    
Expanding near $+\bk_0$ with $\bk = \bk_0 + (\delta k_{\perp}, \delta k_{\parallel})$ gives

\begin{equation}
    \begin{aligned}
        \varepsilon_{\bk} &= \vF\,\delta k_{\perp} + \tfrac{\kappa}{2}\,\delta k_\parallel^2\,,
        \\
        \varepsilon_{\bk+\bq} &= \vF\,\delta k_\perp + \tfrac{\kappa}{2}(\delta k_\parallel+q_\parallel)^2\,.
    \end{aligned}
\end{equation}
The Fermi velocities at both $\bk$ and $\bk+\bq$ is $\vF$. 
Consequently, the linear terms $\vF\,\delta k_\perp$ cancel in the energy difference giving
\begin{equation}
\label{eq:Energy_Conservation_Antipodal}
\varepsilon_{\bk} - \varepsilon_{\bk+\bq} = -\kappa\, q_\parallel \, \delta k_\parallel - \tfrac{\kappa}{2}q_\parallel^2\,.
\end{equation}
The first delta function $\delta(\varepsilon_{\bk})$ fixes 
$\delta k_\perp = -(\kappa/2\vF)\delta k_\parallel^2$, 
and the second delta function then fixes $\kappa q_\parallel \delta k_\parallel = \Omega - \kappa q_\parallel^2/2$ which always has a solution.
Since the energy difference depends only on 
$\delta k_\parallel$, 
the Jacobian of the second delta function is
\begin{equation}
\label{eq:jac_anti}
\mathcal{J}_{\text{ATP}} 
= \bigg|\frac{\partial(\varepsilon_{\bk}-\varepsilon_{\bk+\bq})}{\partial(\delta k_\parallel)}\bigg|
= \kappa\,|q_\parallel|\,.
\end{equation}
This produces a singular factor $1/(\kappa|q_\parallel|)$ after integration over $\delta k_\parallel$.
Including both antipodal points and accounting for the 
Jacobian of the first delta function $1/\vF$ that comes from the $\delta k_\perp$ integral gives us
\begin{equation}
\label{eq:sachdev}
\frac{\ImPi_{\text{ATP}}(\bq, \Omega)}{\Omega}
= \frac{g^2}{2\pi\,\vF\,\kappa\,|q_\parallel|}\,.
\end{equation}
This $1/|q_\parallel|$ gives $\gamma_{\bq}\propto 1/|\bq|$ and $z=3$.
The cancellation of linear energy terms is a direct consequence 
of the parallel velocities at the connected points.
With the linear terms gone, the constraint falls on the quadratic  variable $\delta k_\parallel$, whose Jacobian vanishes at $q_\parallel = 0$.

\subsection{Umklapp Contribution}
\label{subsec:umklapp}

\begin{figure}
	\centering
		\includegraphics[scale=0.6]{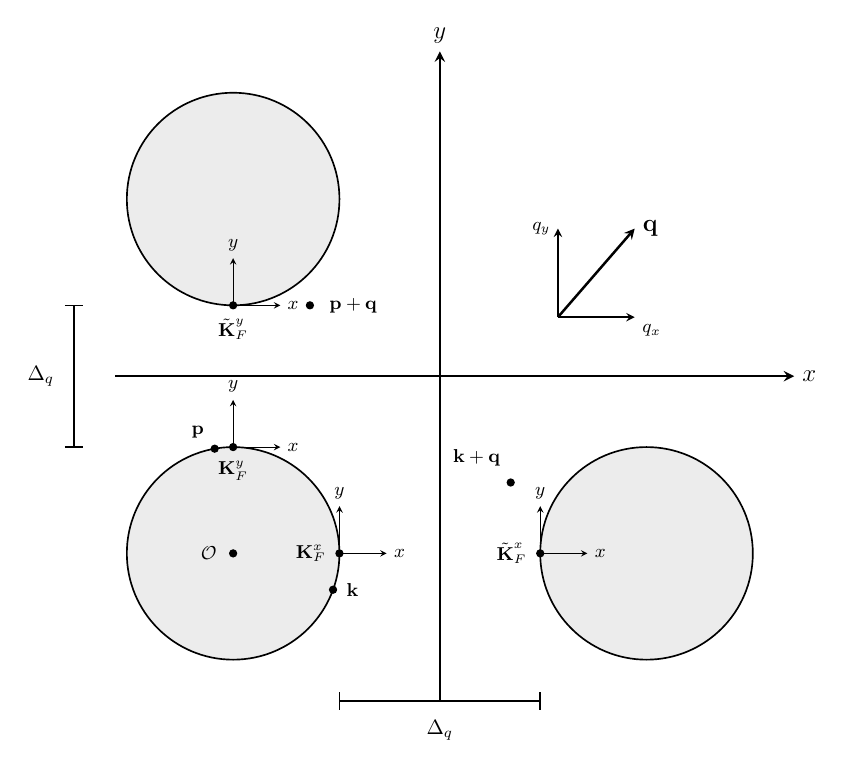}
	\caption{The two umklapp contributions to $\Pi_{\text{U}}$. 
    $\mathcal{O}$ is the center of the original FS.
    Also shown are the two umklapp partners accessible via momentum transfer $\bq$.
    The proportions are greatly exaggerated for visual clarity.
    The analysis in the main text focuses on umklapp along the $x$ direction as the steps for the $y$ direction are exactly the same.}
	\label{fig:Umklapp}
\end{figure}

Without loss of generality we take $\bq = (q_x,q_y)$ with $q_x, q_y \geq 0.$
Consider the FS point $\textbf{K}_F^x = (k_{F},0)$ closest to the BZ boundary 
along $\hat{x}$.  
An electron in the vicinity of $\textbf{K}_F^x$ can scatter by $\bq$ with $q_x\approx \Delta_q$, 
wrapping across the BZ to land in the vicinity of $\textbf{K}_F^x + (\Delta_q, 0) =(k_F + \Delta_q, 0) \equiv \tilde{\textbf{K}}_F^x$.
Due to periodicity we have $\tilde{\textbf{K}}_F^x = (k_F + \Delta_q, 0) = (-k_F, 0)$ in the eyes of the dispersion relation which then gives $\varepsilon_{\textbf{K}_F^x} = \varepsilon_{\tilde{\textbf{K}}_F^x} = 0$.  

The Fermi velocity at $\textbf{K}_F^x$ is $+\vF\,\hat{x}$ and 
at $\tilde{\textbf{K}}_F^x$ it is $-\vF\,\hat{x}$ (by inversion symmetry).  
We set $\bk = \textbf{K}_F^x + (\delta k_x, \delta k_y)$ expressed in a way that allows for a simple Taylor expansion of the dispersion around the zero energy point $\textbf{K}_F^x$ as
\begin{equation}
\label{eq:eps_umkl_1}
    \varepsilon_{\bk} = \vF\, \delta k_x + \tfrac{\kappa}{2}\,\delta k_y^2\,.
\end{equation}

The other choice is more subtle.
We use 
\begin{equation}
    \begin{aligned}
        \bk + \bq 
        &= \textbf{K}_F^x + (\delta k_x + q_x, \delta k_y + q_y) 
        \\
        &= \tilde{\textbf{K}}_F^x + (\delta k_x + q_x - \Delta_q, \delta k_y + q_y) 
        \\
        \implies \bk + \bq 
        &= \tilde{\textbf{K}}_F^x + (\delta k_x + \delta q_x, \delta k_y + q_y),
    \end{aligned}
\end{equation}
where we have implicitly assumed that $\bk + \bq$ is  close to $\tilde{\textbf{K}}_F^x$ which allows for a Taylor expansion of the dispersion about $\tilde{\textbf{K}}_F^x$ as
\begin{equation}
    \tilde{\varepsilon}_{\bk+\bq} 
    = -\vF(\delta k_x + \delta q_x) + \tfrac{\kappa}{2}(\delta k_y+q_y)^2\,.
\label{eq:eps_umkl_2}
\end{equation}
This expansion is justified when $\delta q_x \equiv q_x - \Delta_q \ll \Delta_q$ and the tilde is there to avoid confusion and keep it explicit that the expansion is about the umklapp point.
The correctness of the sign of $-v_F$ can be verified by noting that as either $k_x$ or $q_x$ increase, we go deeper into the umklapp FS thereby lowering the energy and vice versa.
The linearization requires $\vF\neq 0$, which holds for $\Delta_q > 0$ when the FS does not hit the zone boundary.


With these ingredients in place the energy difference now gives
\begin{align}
\varepsilon_{\bk} - \tilde{\varepsilon}_{\bk+\bq} 
= 2\vF\,\delta k_x + \vF\,\delta q_x - \kappa\, \delta k_y\,q_y - \tfrac{\kappa}{2}q_y^2\,.
\label{eq:ediff}
\end{align}
The crucial feature is $2\vF\, \delta k_x$ since the antiparallel velocities cause the linear terms to add rather than cancel.
Substituting $2\vF\delta k_x = -\kappa \, \delta k_y^2$ from $\delta(\varepsilon_{\bk})$ and plugging in $\delta(\Omega + \varepsilon_{\bk} - \tilde{\varepsilon}_{\bk+\bq})$ gives
\begin{equation}
\label{eq:arg}
\Omega + \vF\,\delta q_x - \kappa\, \delta k_y\,q_y - \tfrac{\kappa}{2}q_y^2  - \kappa\, \delta k_y^2= 0
\end{equation}
and we immediately see the $\delta k_y^2$ floating around in the energy conserving constraint compared to Eq.~\eqref{eq:Energy_Conservation_Antipodal}which is the root cause of the difference from the antipodal contributions.

For $q_y = 0$, this reduces to
\begin{equation}
\label{eq:constraint_qy0}
\kappa\,\delta k_y^2 = \Omega + \vF\,\delta q_x \equiv \Omega - \Omega_{0}(q_x)\,,
\end{equation}
where we define the threshold frequency
\begin{equation}
\label{eq:Oth}
\Omega_{0}(q_x) \equiv -\vF\,\delta q_x = \vF(\Delta_q - q_x)\,.
\end{equation}

The constraint~\eqref{eq:constraint_qy0} has real solutions 
$\delta k_y = \pm\sqrt{(\Omega-\Omega_{0})/\kappa}$ 
only when $\Omega > \Omega_{0}$.  
The Jacobian is
\begin{equation}
\label{eq:jac_umkl_b}
\mathcal{J}_{\text{U}} = \bigg|\frac{\partial}{\partial \delta k_y}
\big(\kappa \delta k_y^2 - \Omega + \Omega_{0}\big)\bigg| = 2\kappa\,|\delta k_y|
= 2\sqrt{\kappa(\Omega - \Omega_{0})}\,.
\end{equation}
Summing over the two roots gives us
\begin{equation}
\label{eq:result_qy0}
\frac{\ImPi_{\text{U}}}{\Omega}\bigg|_{\substack{q_y=0}}
= \frac{g^2}{4\pi v_F} \cdot \frac{\Theta\!\big(\Omega - \Omega_{0}\big)\,}{\sqrt{\kappa\,(\Omega - \Omega_0)}}\;
,
\end{equation}
where the $v_F$ comes from $\delta k_x$ integral over $\delta(\varepsilon_{\bk})$.
This is a square-root threshold singularity as 
$\ImPi/\Omega$ diverges as $1/\sqrt{\Omega-\Omega_{0}}$ 
as $\Omega\to\Omega_{0}^+$.

At the umklapp resonance $q_x = \Delta_q \implies \Omega_{0} = 0$, this becomes
\begin{equation}
\label{eq:sqrtOmega}
\ImPi_{\text{U}}(\Delta_q,\Omega) \propto \sqrt{\Omega}\,.
\end{equation}

For $q_y\neq 0$, Eq.~\eqref{eq:arg} gives
\begin{equation}
\kappa\,\delta k_y^2 + \kappa\,\delta k_y\,q_y + \tfrac{\kappa}{2}q_y^2 
= \Omega + \vF\,\delta q_x\,.
\end{equation}
Completing the square gives $\kappa \,\tilde{\delta k_y}^2 = \Omega - \Omega_{0}(\bq)$ 
where $\tilde{\delta k_y} = \delta k_y + q_y/2$ and $\Omega_{0}(\bq) \equiv \vF(\Delta_q - q_x) + \frac{\kappa}{4}q_y^2\, = \tilde{\varepsilon}_{\bq} - \frac{\kappa}{4}q_y^2.$
Eq.~\eqref{eq:result_qy0} maintains the same structure
\begin{equation}
\label{eq:result_general}
\frac{\ImPi_{\text{U}}(\bq, \Omega)}{\Omega}
= \frac{g^2}{4\pi v_F} \cdot\frac{\Theta\!\big(\Omega - \Omega_{0}(\bq)\big)}{\sqrt{\kappa\,(\Omega - \Omega_{0}(\bq))}}\;
\,.
\end{equation}


In the above discussion we considered umklapp bubble in the vicinity of $\textbf{K}_F^x$ with $q_x \approx \Delta_q$.
We also need to account for the contribution from $\textbf{K}_F^y = (0, k_F)$ which requires $q_y \approx \Delta_q$.
Needless to say, the analysis remains exactly the same with the substitution $x \leftrightharpoons y$ throughout.
This means for a given $\bq = (q_x, q_y)$, adding the contributions from $\textbf{K}_F^x$ and $\textbf{K}_F^y$ 
\begin{equation}
\label{eq:Umklapp_Pi_Final}
\frac{\ImPi_{\text{U}}(\bq, \Omega)}{\Omega}
= \frac{g^2}{4\pi v_F} \Bigg[
\frac{\Theta\!\big(\Omega - \Omega_{0}(\bq)\big)}{\sqrt{\kappa\,(\Omega - \Omega_{0}(\bq))}}\;
+
\frac{\Theta\!\big(\Omega - \Omega_{0}(\bar{\bq})\big)}{\sqrt{\kappa\,(\Omega - \Omega_{0}(\bar{\bq}))}}\;
\Bigg]
\, ,
\end{equation}
with $\bar{\bq} = (q_y, q_x)$.

\subsection{Extension to Three Dimensions}
\label{sec:3D}

The key features of the calculation generalize to 3D 
with modifications to the dimensionality of the transverse integration.
In 3D, the FS is a two-dimensional surface, 
and the patch expansion near an umklapp point involves 
two transverse momenta $(\delta k_y, \delta k_z)$ rather than one ($\delta k_y$) in 2D.  
After the FS constraint fixes the longitudinal momentum $\delta k_x$, 
the energy conservation becomes a constraint on $(\delta k_y, \delta k_z)$ as
\begin{equation}
\kappa_1 \delta k_y^2 + \kappa_2 \delta k_z^2 = \Omega - \Omega_{0}(\bq)\,,
\end{equation}
where $\kappa_1$, $\kappa_2$ are the two principal curvatures 
at the umklapp point.  
This is an ellipse in $(\delta k_y, \delta k_z)$ space and gives
\begin{equation}
\int d \delta k_y \, d \delta k_z\;\delta(\kappa_1 \delta k_y^2 + \kappa_2 \delta k_z^2 - E)
= \frac{\pi}{\sqrt{\kappa_1\kappa_2}}\,\Theta(E)\,.
\end{equation}

Therefore, in 3D the umklapp contribution to $\ImPi/\Omega$ 
has a step function threshold and no square-root singularity
\begin{equation}
\ImPi^{\text{3D}}_{\text{U}}/\Omega \sim \Theta(\Omega - \Omega_{0})\,.
\end{equation}

This implies that for critical heavy fermion systems, the umklapp self energy $\Pi_{\text{U}}$ plays no serious role. 
This conclusion does not change even when $\ImPi^{\text{3D}}_{\text{U}}$ is evaluated using dressed fermion propagators as can be seen at the end of App.[\ref{app:A}].

\section{Implications For Transport Near Criticality}

\label{sec:Implications}

As was noted in \cite{Wang_Berg, Lee_umklapp}, near Ising-nematic criticality $r \rightarrow 0,$ the minimum umklapp momenta $\Delta_q$ gives a dynamically generated temperature scale $T_U$ which decides the crossover from $\rho(T) \sim T^2$ to $\rho(T) \sim T$ as $T$ increases from $0$.
After incorporating $\Pi_{\rm{U}}$ and the boson thermal mass, the fully dressed boson propagator written in terms of real frequency $\Omega$ for the values of $\bq$ when $\Omega_0(\bq)=0$ is
\begin{equation}
    D^{-1}(\bq, \Omega) = q^2 + aT - \eta\frac{i \Omega}{q} - i \lambda  \sqrt{\Omega},
\end{equation}

For $q \sim \Delta_q$ this furnishes additional energy scales.
Balancing $\eta \Omega/q$ with bare $q^2$ gives the $z=3$ crossover scale $T_3 \sim \eta^{-1} \Delta_q^3$ \cite{Wang_Berg, Lee_umklapp}.
Comparing $\lambda\sqrt{\Omega}$ with the bare kinetic term $q^2$ gives $T_4 \sim \lambda^{-2} \Delta_q^4$.
Balancing the boson thermal mass with the bare kinetic term gives $T_m \sim a^{-1}\Delta_q^2$
And finally comparing the two pieces of boson self energy $\eta \Omega/q$ vs $\lambda \sqrt{\Omega}$ gives $T_2 \sim \lambda^2 \Delta_q^2 \eta^{-2}$.
\cite{Wang_Berg, Lee_umklapp} only have $T_3$ as their relevant temperature scale giving $T_U \sim T_3$ and there are no analogues of $T_4, T_2$ and $T_m$.

Making use of $\Delta_q$ as the source of a small parameter furnishes us a hierarchy of scales
\begin{equation}
    T_4 \ll T_3 \ll T_2 \sim T_m.
\end{equation}

Now the question is under what circumstances is the above relative $T$ scale meaningful and consistent and when can we expect the reduced temperature scale $T_4$ to act as the crossover scale $T_U$?



\subsection{Memory Matrix Transport}

In \cite{Wang_Berg}, the authors evaluated umklapp transport resistivity at criticality and found a crossover temperature $T_U \sim T_3$ ($T_0$ in their notation) which a priori demonstrates that $T_4$ plays no role.
Here we discuss why this disagreement comes about.

They parametrized the boson propagator as 
$D^{-1} = r_{\bq} + \gamma_{\bq}|\Omega|$ and suspected that umklapp could modify $\gamma_{\bq}$ and thereby alter $z=3$.
Although $\gamma_{\bq} \sim 1/q$ can be calculated for spherical dispersions, they set $\gamma_{\bq} \sim \sum_{\bk} \delta(\varepsilon_{\bk})\delta(\varepsilon_{\bk + \bq})$ as it allowed them to numerically capture the non-universal effects by performing the summation for general lattice dispersion $\varepsilon_{\bk}$.
A direct comparison with Eq.\eqref{eq:ImPi_T0} gives $\gamma_{\bq} \sim \sum_{\bk} \delta(\varepsilon_{\bk})\delta(\varepsilon_{\bk + \bq}- \Omega)$.
Ignoring $\Omega$ inside the second delta function is unproblematic for capturing the antipodal $\gamma_{\bq} \sim 1/q$ contribution as can be seen from the discussion following Eq.\eqref{eq:Energy_Conservation_Antipodal}.

Our result shows that the umklapp correction does \textit{not} 
take the form of an $\Omega$ independent $\gamma_{\bq}$.  
Instead, it introduces a qualitatively new term 
$\sqrt{\Omega}$ in the boson self-energy 
at $q \sim \Delta_q$ which explicitly requires keeping $\Omega$ in $\delta(\varepsilon_{\bk+\bq} - \Omega)$ as can be seen from the discussion around Eq.\eqref{eq:constraint_qy0}.  
This term is not captured by the simple $\gamma_{\bq}|\Omega|$ 
ansatz of \cite{Wang_Berg} and does not reflect in the transport properties they evaluate.
Redoing the memory matrix analysis with $\gamma_{\bq} \sim \sum_{\bk} \delta(\varepsilon_{\bk})\delta(\varepsilon_{\bk + \bq}- \Omega)$ would hopefully capture the effects of $\Pi_{\rm{U}}$ at the level of transport, but this lies much outside the scope of this work.

\begin{figure}
	\centering
		\includegraphics[scale=0.35]{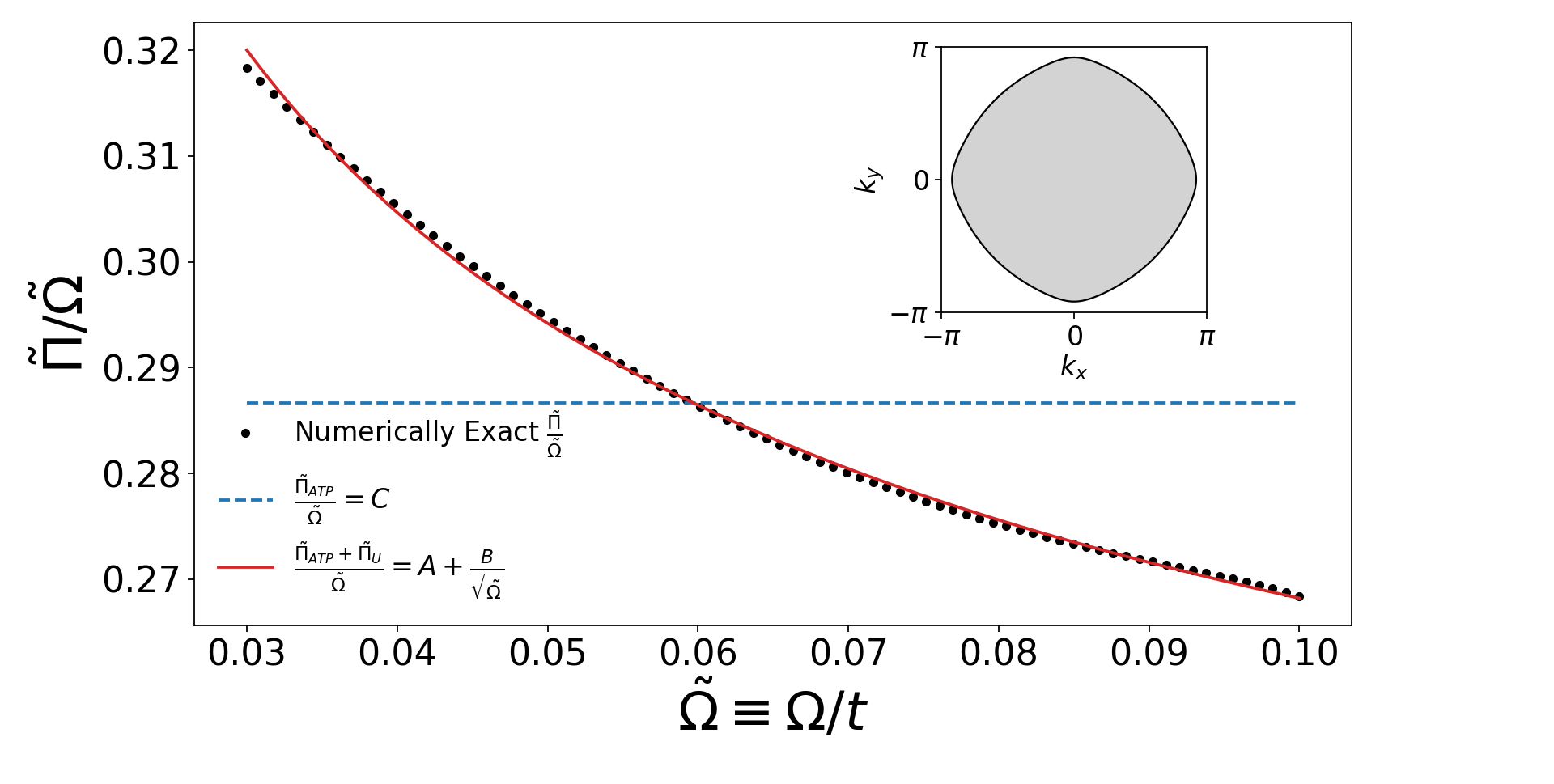}
	\caption{The frequency divided boson self energy as a function of boson frequency for fixed $\bq = (\Delta_q, 0)$.
    The tildes ensure that everything is dimensionless.
    The dispersion is from \cite{Wang_Berg}  $\varepsilon_{\textbf{k}} = -2t (\cos(k_x) + \cos(k_y)) -4t' \cos(k_x) \cos(k_y) - \mu$.
    We choose $(t, t', \mu) = (1.0, \, 0.3, \, 1.1)$ where the accompanying FS is in the inset.
    For values of $\tilde{\Omega}$ smaller than the ones shown in the plot, deviations kick in due to the numerical implementation of the Dirac delta function which can be remedied by a smaller regulator at the expense of longer runtime.
    For values of $\tilde{\Omega}$ larger than the ones shown in the plot we again see deviations but those are due to the breakdown of the patch approximation and the $\tilde{\Omega}^{-1/2}$ behavior is not expected for such high frequencies.
    Within the practical and theoretical bound we see a nice agreement of the exact result with the calculation of $\Pi_{\rm{U}}$ performed in this work.
    }
	\label{fig:Self_Energy_Curve_Fit}
\end{figure}

As a sanity check for the validity of the calculations performed till now, we plot the full $\Pi/ \Omega \propto \sum_{\bk} \delta(\varepsilon_{\bk})\delta(\varepsilon_{\bk + \bq}- \Omega)$ from Eq.\eqref{eq:ImPi_T0} as a function of $\Omega$ for the dispersion from \cite{Wang_Berg} numerically and compare it with $\Pi = \Pi_{\text{ATP}} + \Pi_{\rm{U}}$ from analytical calculations shown above.
We set $\bq = (\Delta_q, 0)$ which ensures that $\Omega_0(\bq) = 0$. 
Now for a fixed $\bq$, purely from the antipodal contribution we expect $\Pi_{\text{ATP}}/\Omega$ to be a constant in $\Omega$ whereas  $\Pi = \Pi_{\text{ATP}} + \Pi_{\rm{U}}$ would give rise to an additional $1/\sqrt{\Omega}$ from Eq.\eqref{eq:result_qy0} whose effects can be seen clearly in Fig.[\ref{fig:Self_Energy_Curve_Fit}].

\subsection{Bottlenecks}
\label{subsec:Bottlenecks}
In evaluating $\Pi(\bq, \Omega)$ we've only used free fermion propagators.
The dressed boson propagator gives rise to a fermion self energy $\Sigma(i \omega)$ which for non-Fermi liquids (NFL) gives rise to a temperature scale $T_{\rm{NFL}}$ below which $\Sigma(i \omega)$ dominates over the bare $i \omega$ of the fermion propagator and the quasi-particle picture is invalidated \cite{Sachdev_QPT, Sachdev_Phases}.

An important point to note is that due to the momentum activation threshold, $\Pi_{\rm{U}}$ does not modify fermion self energy in any meaningful way.
When evaluating $\Sigma(i \omega)$ using the fully dressed $D(\bq, i \Omega)$ including both, $\Pi_{\rm{ATP}}$ and $\Pi_{\rm{U}}$, the effects of $\Pi_{\rm{U}}$ kick in only when $q \sim \Delta_q$.
In this regime the boson propagator is effectively gapped with a mass $\sim \Delta_q^2$ which at worst only gives Fermi liquid like self energy corrections with pre-factors that depend on $\Delta_q$ but it never really competes with $\Sigma(i \omega)$ and the dressed NFL propagator remains completely intact.
Now re-evaluating $\Pi_{\rm{ATP}}$ using the dressed fermion propagator gives back the same $\Omega/q$ result \cite{Sachdev_Phases}.

Re-evaluating $\Pi_{\rm{U}}$ using the dressed NFL fermion propagator weakens the contribution from 
\begin{equation}
    \Pi_{\rm{U}} \sim \sqrt{\Omega} \longrightarrow \Omega^{2/3}.
\end{equation}
A hand-wavy way of seeing this is as follows.
Ignoring $\Omega_0$ for the moment, we saw that $\Pi_{\rm{U}}/\Omega \propto 1/\sqrt{\Omega}$.
In Eq.~\eqref{eq:result_general}, the square root comes from the integral over $\delta k_y^2$ and the $\Omega$ inside the square root comes from the $i\omega$ kinetic term in the free fermion propagator.
In the NFL regime we can ignore the $i \omega$ term and worry only about $\Sigma(i\omega)$ in the dressed fermion propagator to obtain 
\begin{equation}
    \Pi_{\rm{U}} \propto \frac{\Omega}{\sqrt{\Sigma(i\Omega)}}
\end{equation}
This argument is fairly schematic, does not hold under technical scrutiny and has only been presented to show transparently why the correct answer (derived in App.[\ref{app:A}]) looks the way it does.

Now for $(d, z) = (2, 3)$ criticality, $\Sigma(i\Omega) \sim \Omega^{2/3}$ which immediately gives $\Pi_{\rm{U}} \sim \Omega^{2/3}$. 
Even at $q\sim \Delta_q$ and very low $\Omega$ when $\Omega^{2/3}$ dominates over $\Omega/q$, $\Omega^{2/3} \sim \Delta_q^2$ still gives $\Omega \sim \Delta_q^3$ so $z=3$ criticality remains intact even at higher values of $q$ for $T \rightarrow 0$.


\subsection{Consistency Requirements}

For the lowered temperature $T_4$ to exist robustly we need the following condition
\begin{equation}
    T_{\rm{NFL}} \ll T_4,
\end{equation}
where $T_{\rm{NFL}}$ is the temperature scale above which the bare $i \omega$ dominates over $\Sigma(i \omega)$ in the dressed fermion propagator.
Working above $T_{\rm{NFL}}$ justifies the use of free fermion propagators leading to $\Pi_{\rm{U}} \sim \sqrt{\Omega}$ which then gives rise to $T_4$.
If $T_4 \ll T_{\rm{NFL}}$ then $T_4$ and $T_2$ never shows up since below $T_{\rm{NFL}}$ we have $\Pi_{\rm{U}} \sim \Omega^{2/3}$ which gives rise to the exact same $T_3$ as was obtained from $\Omega/q$ up to numerical pre-factors.

We use the fact that for spherical FS, $v_F = k_F/m$, $\kappa = m^{-1}$ 
and $\varepsilon_F \sim k_F^2/m$ where $k_F$ is the Fermi momenta and $m$ is the fermion mass.
This gives $T_4 \sim (\Delta_q/k_F)^4 \, \varepsilon_F^3 g^{-4} $ and $T_3 \sim (\Delta_q/k_F)^3 \, \varepsilon_F^2 g^{-2} $ since $\lambda \sim g^2m^{3/2} k_F^{-1}$ and $\eta \sim  g^2m^{2} k_F^{-1}$ as can be seen from Eq.\eqref{eq:Umklapp_Pi_Final} and $T_2 \sim (\Delta_q/k_F)^2 \varepsilon_F$ having ignored factors of $2$ and $\pi$ etc.

Now we have to check when exactly is $ T_{\rm{NFL}} \ll T_4$.

\subsubsection{Yukawa-SYK model}

For Y-SYK model $T_{\rm{NFL}} \sim g^4\varepsilon_F^{-1}$ \cite{Sachdev_Phases} and the requirement $ T_{\rm{NFL}} \ll T_4 \implies   g^2 \varepsilon_F^{-1}  \ll \Delta_q/k_F$ is completely inconsistent with (and in fact the exact opposite of) $T_4 \ll T_3 \implies  \Delta_q/k_F \ll g^2 \varepsilon_F^{-1}$ so in this large-$N$ limit, the following condition does not hold consistently
\begin{equation}
     T_{\rm{NFL}} \ll T_4 \ll T_3.
\end{equation}

When $\Delta_q/k_F \ll g^2 \varepsilon_F^{-1}$, the condition that actually holds is
\begin{equation}
    T_4 \ll T_3 \ll T_{\rm{NFL}}.
\end{equation}
But just because a condition is consistent doesn't mean it's physically meaningful.
As discussed up above, below $T_{\rm{NFL}}$, we have $\Pi_{\rm{U}} \sim \Omega^{2/3}$ and $\Pi_{\rm{ATP}} \sim \Omega/q$ both of which give $T_3$ as the physically relevant scale and $T_4$ does not show up in this case so the possibility of $T_U \sim T_4$ can be rejected immediately.

The other consistent possibility comes from $\Delta_q/k_F \gg g^2 \varepsilon_F^{-1}$ giving
\begin{equation}
    T_{\rm{NFL}} \ll T_2 \ll T_3 \ll T_4.
\end{equation}
But in this case, $g^2 \varepsilon_F^{-1}$ acts as the small parameter making  $T_2$ the smallest scale and in fact above $T_2$, $\Pi_{\rm{ATP}}$ strongly dominates over $\Pi_{\rm{U}}$ so we would still expect $T_U \sim T_3$ for this case.

\begin{figure}
	\centering
		\includegraphics[scale=0.3]{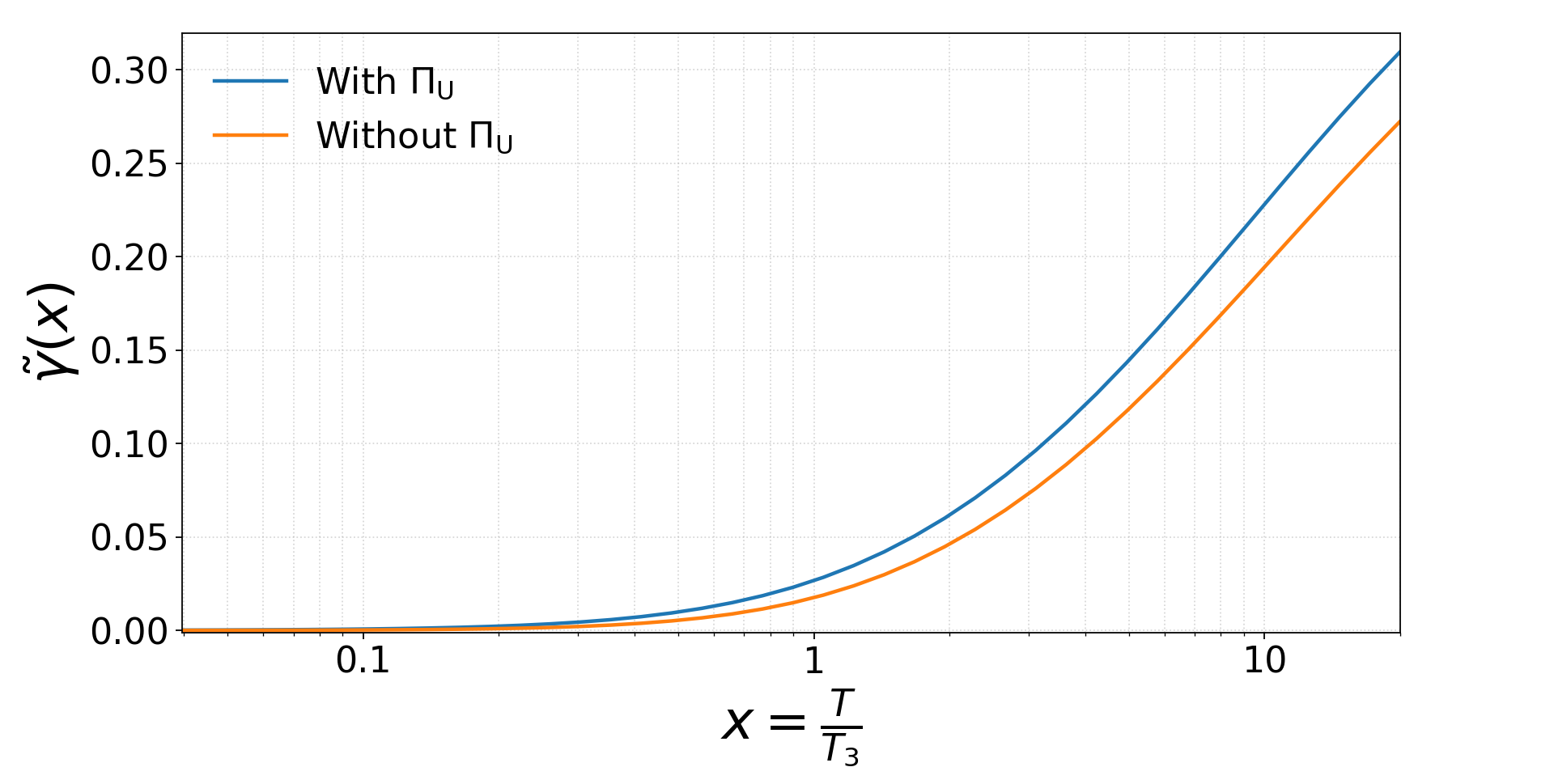}
	\caption{The scattering rate of the electron at $\textbf{K}_F^x$ in terms of dimensionless temperature for the Y-SYK model above $T_{\rm{NFL}}$.
    The boson thermal mass has been included here with $T_m = T_4$ set by hand.
    Here $\Delta_q/k_F = \alpha \, g^2 \varepsilon_F^{-1}$ with $\alpha = 10$ which sets $T_2 = \alpha^{-1} T_3 \ll T_3  \ll T_4 = \alpha T_3$.
    The three tick labels shown on the x-axis correspond to $T = T_2, \; T_3$ and $T_4$ respectively.
    The effect of $\Pi_{\rm{U}}$ is only minute and quantitative}
	\label{fig:Scattering_rate_SYK}
\end{figure}

\subsubsection{Naive large-$N$ limit}

\begin{figure}
	\centering
		\includegraphics[scale=0.3]{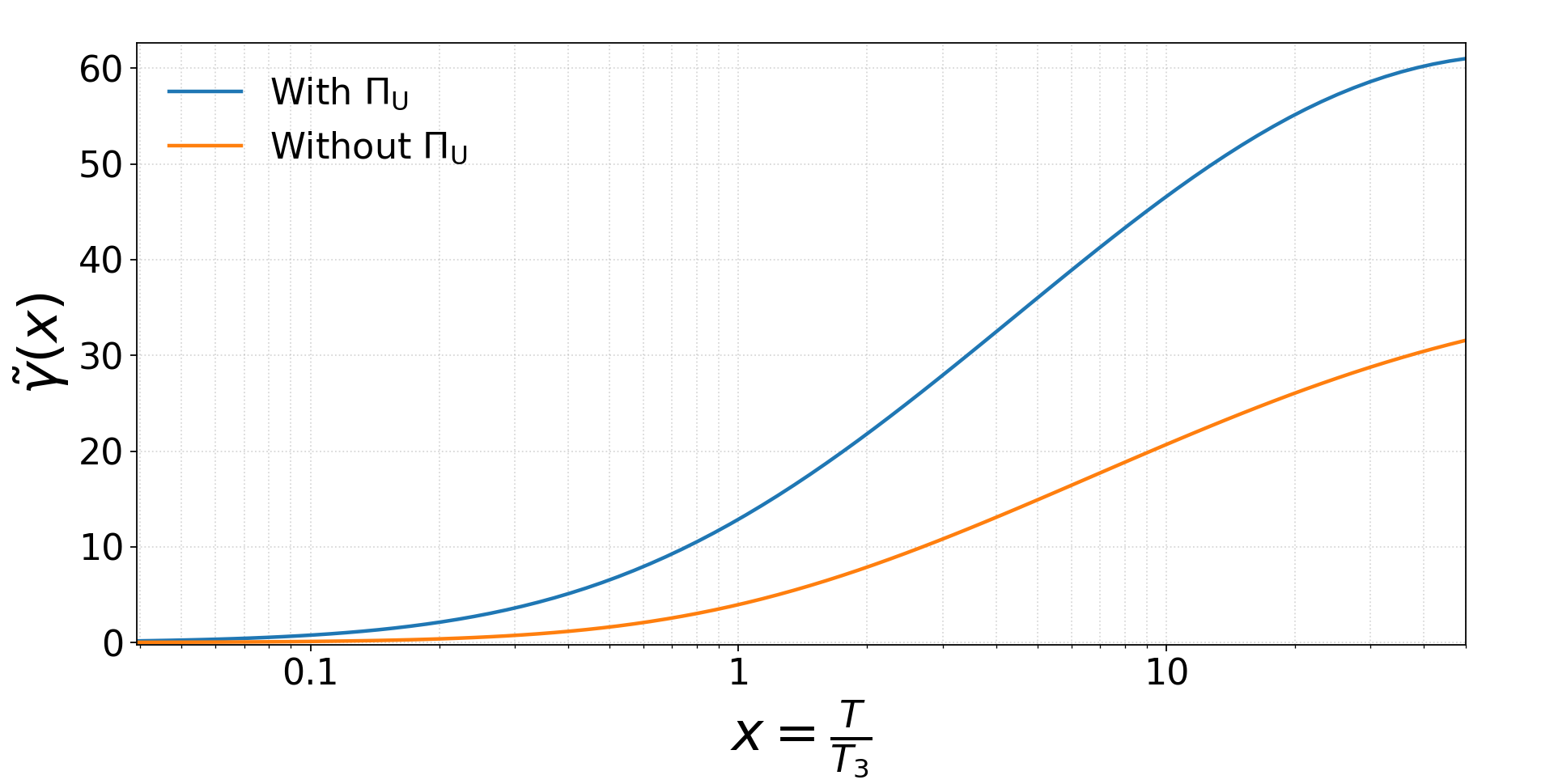}
	\caption{The same plot for the naive large-$N$ limit.
    The boson thermal mass has been included in both cases with $T_2 = T_m$ for simplicity which leads to $\sqrt{x}$ saturation at large $x$.
    Here $\Delta_q/k_F  = 0.1 \, g^2 \varepsilon_F^{-1}$ with $\alpha = 0.1$ which sets $T_4 = \alpha T_3 \ll T_3 \ll T_2 = T_m = \alpha^{-1} T_3$.
    The three tick labels shown on the x-axis correspond to $T = T_4, \; T_3$ and $T_2$ respectively.
    $\Pi_{\rm{U}}$ leads to a substantial enhancement of the scattering rate beyond $T_4$.}
	\label{fig:Scattering_rate_Naive}
\end{figure}

In the naive large-$N$ limit with $N$ species of fermions and a single bosonic field we have $T_{\rm{NFL}} \sim g^4\varepsilon_F^{-1} N^{-3}$ \cite{Wang_Berg} which is adequately suppressed for large enough $N$.
Now requiring  $T_{\rm{NFL}} \ll T_4$ immediately gives  $N^{-3/4} g^2 \varepsilon_F^{-1}  \ll \Delta_q/k_F$  which can be satisfied consistently with $T_4 \ll T_3 \implies \Delta_q/k_F \ll g^2 \varepsilon_F^{-1}$.
This is precisely the large-$N$ limit and temperature regime studied in \cite{Wang_Berg}.
In this case the following condition holds consistently
\begin{equation}
     T_{\rm{NFL}} \ll T_4 \ll T_3 \ll T_2,
\end{equation}
which suggests the possibility that $T_U \sim T_4$ but  this needs to be confirmed using a more careful transport analysis.
It is called the naive large-$N$ limit due to the technical problems that occur for $T \ll T_{\rm{NFL}}$ as the $N^{-1}$ expansion becomes ill defined \cite{Sachdev_Chowdhury_Review, SSLee_Patch} and other methods are needed as a remedy \cite{Mandal_1, Dim_Reg}.

\subsubsection{Matrix large-$N$ limit}

In this case the scalar field comes with $N^2$ matrix valued components and gives rise to $T_{\rm{NFL}} \sim g^4\varepsilon_F^{-1} N$ \cite{Matrix_Raghu}.
Since both $\Pi_{\rm{ATP}}$ and $\Pi_{\rm{U}}$ are $\mathcal{O}(N^{-1})$ for this model we immediately get $T_3 \sim \mathcal{O}(N), \; T_4 \sim \mathcal{O}(N^2)$ and $T_2 \sim \mathcal{O}(1)$.

Although this case is the most messy since we have 3 different small parameters $N^{-1}, \Delta_q/k_F$ and $g^2/\varepsilon_F$, for large enough $N$ the lowest $T$ scale is always $T_2$ above which $\Pi_{\rm{ATP}}$ completely dominates $\Pi_{\rm{U}}$ and again we expect nothing very interesting.

\section{Scattering Rate}
\label{sec:Scattering_Rate}

For SYK and naive large-$N$ models for temperatures above $T_{\rm{NFL}}$, we reanalyze the scattering rate in \cite{Lee_umklapp} with two modifications.
We include the boson thermal mass $aT$ and include $\Pi_{\text{U}}$ in the boson propagator.
For simplicity, we focus on the scattering rate of the electron that originally sits at $\textbf{K}_F^x$ to scatter and land in the vicinity of $\tilde{\textbf{K}}_F^x$ right on the umklapp FS from Fig.[\ref{fig:Umklapp}] as was the case at the start of \cite{Lee_umklapp}.
Since the scattering rate obtained in \cite{Lee_umklapp} didn't involve the boson mass, for $T > T_3$ an indefinite linear in $T$ scattering rate was obtained.
It was at the level of quantum Boltzmann equation along with the relaxation time approximation that it became clear that there is another scale $T_1 \propto \gamma_0 \Delta_q$ set by the potential disorder $\gamma_0$ above which the linear in $T$ \textit{resistivity} saturates to a $\sqrt{T}$ dependence.

The exact same simple scattering rate when evaluated including the boson thermal mass gives
\begin{equation}
     \tilde{\gamma}(T) \approx T \int_{-\infty}^{\infty}  \frac{d \phi}{q^2_\phi + aT} \propto \frac{T}{\sqrt{\Delta_q^2 + aT}},
\end{equation}
which immediately gives a scale $T_m \sim a^{-1}\Delta_q^2$ above which the scattering rate becomes $\sqrt{T}$.
In the accompanying plots we arbitrarily set $T_m$ as exactly the same as the largest relevant $T$ scale. 

In Fig.[\ref{fig:Scattering_rate_SYK}], for $T_2 < T < T_3,$ the effects of $\Pi_{\rm{U}}$ seem extremely benign and only quantitative as discussed earlier for Y-SYK model.
The physically appropriate choice would be $\Delta_q/k_F = 0.1, \; g^2 \varepsilon_F = 0.01$ but in the evaluation its the relative ratios that matter so we set $\Delta_q/k_F = 10, \; g^2 \varepsilon_F^{-1} = 1.0$

In Fig.[\ref{fig:Scattering_rate_Naive}] for the naive large-$N$ model, we can see that due to the presence of $\Pi_{\rm{U}},$ the scattering rate gets enhanced at a much lower $T\sim T_4$ compared to what's expected without $\Pi_{\rm{U}}$ where we see an upturn only around $T_3$.
At higher temperatures, both scattering rates saturate due to the boson thermal mass.
The details regarding the evaluation of these plots can be found in App.[\ref{app:B}].

\section{Conclusion}
\label{sec:summary}

We computed the umklapp contribution to the  particle--hole polarization bubble for $d=2$ Ising-nematic metal whose Fermi surface approaches the Brillouin zone boundary. 
The key result is that umklapp processes do not simply renormalize the Landau damping coefficient $\gamma_q \sim 1/q$ as was previously posited~\cite{Wang_Berg}. 
Instead, they introduce a qualitatively distinct contribution $\Pi_{\mathrm{U}} \propto \Omega^\alpha$, where $\alpha = 2/3$ for $T \ll T_{\rm{NFL}}$ and $\alpha = 1/2$ for $T \gg T_{\rm{NFL}}$ at the minimum umklapp momentum $q \sim \Delta_q$, separated from $q=0$ by a frequency threshold $\Omega_0(\mathbf{q})$. 
For $T \ll T_{\rm{NFL}}$, both $\Omega^{2/3}$ and $\Omega/q$ give $z=3$.
This conclusion does not hold for model deformations with generalized boson dispersion \cite{Senthil_Controlled_Expansion} $q^2 \rightarrow q^{1+\epsilon}$ where additional competing scales are generated. The schematic formula (in the non-Fermi liquid regime)
\begin{equation}
    \Pi_{\rm{U}}(\Omega) \propto \frac{\Omega}{\sqrt{\Sigma(i\Omega)}},
\end{equation}
immediately gives a first order insight into the potential importance of $\Pi_{\rm{U}}$.

The impact of $\Pi_{\mathrm{U}}$ is strongly controlled by the nature of the large-$N$ limit.
For Yukawa-SYK and matrix large-$N$ models, within the scope of this work, the effects of $\Pi_{\mathrm{U}}$ seem inconsequential.
For the naive large-$N$ model there is a possibility for a lower temperature $T_4$ acting as the crossover scale for the activation of critical umklapp transport compared to $T_3$ as is previously expected.
We've shown this to be the case at the level of simple scattering rate but whether this actually modifies $T_U$ depends on a transport calculation that goes beyond the present work.

For physically realistic systems, $N \sim \mathcal{O}(1)$ so the energy scale estimates made for Y-SYK model seem most realistic as there we do not have pre-factors of $N$ artificially suppressing or inflating any $T$ scale which leads us to believe that for realistic systems the effects of $\Pi_{\rm{U}}$ would be benign.



In $d=3$, the additional transverse momentum integration removes the square-root dependence to $\Pi_{\mathrm{U}}^{3D} \sim \Omega$, which does not compete with the antipodal $\Omega/q$ damping. 
Consequently, despite the original physical motivation for the current work \cite{JvD_PAM1, JvD_PAM2, Paul_pepin_2, Paul_Pepin_3, Pepin_Paul_1}, we find that the umklapp correction plays no significant role for three-dimensional critical systems such as heavy-fermion metals near a Kondo-breakdown transition.

\section{Acknowledgments}
\label{sec:acknowledgments}

We thank Stefan Kehrein for important feedback on the manuscript.

\begin{widetext}

\appendix

\section{Umlapp self energy at low temperatures}
\label{app:A}

We start with the self consistent patch calculation of the antipodal piece as it makes the subsequent calculations more illuminating.
Starting with $\bq = (0, q_\parallel)$,
\begin{equation}
    \begin{aligned}
\varepsilon_{\bk} &= \vF\,\delta k_{\perp} + \tfrac{\kappa}{2}\,\delta k_\parallel^2
\\
\varepsilon_{\bk+\bq} &= \vF\,\delta k_\perp + \tfrac{\kappa}{2}(\delta k_\parallel+q_\parallel)^2\,,
\end{aligned}
\end{equation}
we can write down the antipodal part of the boson self energy bubble as \cite{Sachdev_Phases}
\begin{equation}
    \begin{aligned}
\Pi_{\rm{ATP}}\left(\bq, i \Omega_m\right)= & -2 g^2 T \sum_{\omega_n} \int \frac{d^2 k}{4 \pi^2} \frac{1}{\left( W(i \omega_n)-v_F \delta k_\perp-\frac{\kappa}{2} \delta k_\parallel^2 \right)}
\cdot \frac{1}{\left(W\left(i\omega_n + i\Omega_m\right)-v_F k_\perp-\frac{\kappa}{2}\left(\delta k_\parallel+q_\parallel\right)^2 \right)}
\end{aligned}
\end{equation}
where  we use a placeholder $W(i\omega) \equiv  i\omega - \Sigma(i\omega)$ to avoid clutter.
Performing the integral over $\delta k_\perp$ gives
\begin{equation}
    \begin{aligned}
\Pi_{\rm{ATP}}\left(\boldsymbol{q}, i \Omega_m\right)= & \frac{-i g^2 T}{v_F} \sum_{\omega_n} \int \frac{d k_y}{(2 \pi)} 
\frac{\operatorname{sgn}\left(\omega_n+\Omega_m\right)-\operatorname{sgn}\left(\omega_n\right)}{W\left(i\omega_n + i\Omega_m\right) - W\left(i\omega_n\right)-\frac{\kappa}{2} q_\parallel^2 -\kappa q_\parallel k_\parallel} .
\end{aligned}
\end{equation}
and performing the contour integral over $\delta k_\parallel$ gives \cite{Sachdev_Phases}
\begin{equation}
    \begin{aligned}
\Pi_{\rm{ATP}}\left(\bq, i \Omega_m\right) & =\frac{g^2 T}{2 \kappa v_F\left|q_\parallel\right|} \sum_{\omega_n} \operatorname{sgn}\left(\Omega_m\right)\left[\operatorname{sgn}\left(\omega_n+\Omega_m\right)-\operatorname{sgn}\left(\omega_n\right)\right] 
= -\frac{g^2\left|\Omega_m\right|}{2 \pi \kappa v_F\left|q_\parallel\right|}.
\end{aligned}
\end{equation}

For the umklapp contribution the fermion dispersion that acts as the starting point is
\begin{equation}
    \begin{aligned}
\varepsilon_{\bk} &=  \vF\, \delta k_x + \tfrac{\kappa}{2}\,\delta k_y^2\,
\\
\tilde{\varepsilon}_{\bk+\bq} 
 &= -\vF(\delta k_x + \delta q_x) + \tfrac{\kappa}{2}(\delta k_y+q_y)^2\,.
\end{aligned}
\end{equation}
As explained in Sec.[\ref{subsec:Bottlenecks}], despite this umklapp contribution, the fermion self energy $\Sigma(i\omega_n)$ remains unchanged.
For the $d=2$ Yukawa-SYK model $\Sigma(i \omega) = -iB \rm{sgn}(\omega) |\omega|^{2/3}$ with $\sqrt{3}B= g^{4/3}\kappa^{1/3} (v_F/2\pi)^{-2/3}$ \cite{Sachdev_Phases}.
The umklapp particle--hole bubble in Matsubara frequencies is
\begin{equation}\label{eq:Pi_start}
    \Pi_{\rm{U}}(\bq,i\Omega_m) = -g^2 T\sum_{\omega_n}\int\frac{d^2k}{(2\pi)^2}\;\frac{1}{\big(W(i\omega_n) - \vF\, \delta k_x - \tfrac{\kappa}{2}\,\delta k_y^2 )
    \big(W(i(\omega_n{+}\Omega_m)) + v_F(\delta k_x + \delta q_x) - \frac{\kappa}{2}(\delta k_y+q_y)^2\big)}\,.
\end{equation}
The first factor comes from the propagator at the patch near $\vb{K}_F^x$, the second from the umklapp partner $\tilde{\vb{K}}_F^x$ from Fig.[\ref{fig:Umklapp}]. 

The integrand takes the form
\begin{equation}
    \frac{1}{(\alpha - v_F\delta k_x)(\beta + v_F\delta k_x)} = \frac{-1}{v_F^2}\cdot\frac{1}{(\delta k_x - \alpha/v_F)(\delta k_x + \beta/v_F)}\, ,
\end{equation}
where the two poles in the complex $\delta k_x$ plane
\begin{align}
    P_1 &= \frac{\alpha}{v_F}\,,\qquad \im(P_1) = \frac{\im\,W(i\omega_n)}{v_F} \propto \sgn(\omega_n)\,,\label{eq:P1}\\
    P_2 &= -\frac{\beta}{v_F}\,,\qquad \im(P_2) = -\frac{\im\,W(i(\omega_n{+}\Omega_m))}{v_F} \propto -\sgn(\omega_n{+}\Omega_m)\,.\label{eq:P2}
\end{align}
are defined in terms of
\begin{equation}
\label{eq:alpha}
    \begin{aligned}
    \alpha &\equiv W(i\omega_n) - \frac{\kappa}{2}\delta k_y^2\,,
    \\
    \beta &\equiv W(i(\omega_n{+}\Omega_m)) + v_F\delta q_x - \frac{\kappa}{2}(\delta k_y + q_y)^2\,.
\end{aligned}
\end{equation}

For $\omega_n > 0$, we pick up the residue at $P_1$:
\begin{equation}
    \oint \frac{d\,\delta k_x}{2\pi}\;\frac{-1}{v_F^2}\cdot\frac{1}{(\delta k_x - P_1)(\delta k_x - P_2)} = \frac{-1}{v_F^2}\cdot\frac{i}{P_1 + \beta/v_F} = \frac{-i}{v_F(\alpha+\beta)}\,.
\end{equation}
For $\omega_n < -\Omega_m$, we pick up the residue at $P_2 = -\beta/v_F$:
\begin{equation}
    \frac{-1}{v_F^2}\cdot\frac{i}{-\beta/v_F - P_1} = \frac{-1}{v_F^2}\cdot\frac{i}{-(\alpha+\beta)/v_F} = \frac{+i}{v_F(\alpha+\beta)}\,.
\end{equation}
Combining both sectors, the $\delta k_x$ integral gives
\begin{equation}\label{eq:dkx_result}
   \int\frac{d\,\delta k_x}{2\pi}\;G_1\cdot G_2  
   = \frac{-i\,\big[\sgn(\omega_n) + \sgn(\omega_n{+}\Omega_m) \big]}{2v_F(\alpha+\beta)}\;\cdot
\end{equation}
The factor $\frac{1}{2}[\sgn(\omega_n{+}\Omega_m) + \sgn(\omega_n)]$ equals $\pm 1$ when both signs agree, and zero otherwise.
The denominator can be written as
\begin{equation}
\label{eq:apb}
    \begin{aligned}
        \alpha + \beta 
        &= W(i\omega_n) + W(i(\omega_n{+}\Omega_m)) + v_F\delta q_x - \kappa\delta k_y^2 - \kappa\delta k_y q_y - \frac{\kappa}{2}q_y^2\, 
        \\
        &= W(i\omega_n) + W(i(\omega_n{+}\Omega_m)) - \Omega_0(\bq) - \kappa \delta \tilde{k}_y^2,
    \end{aligned}
\end{equation}
which gives us
\begin{equation}
    \Pi_{\rm{U}}(\bq, i\Omega_m) 
    = \frac{ig^2 T}{2v_F}
    \sum_{\omega_n}\big[\sgn(\omega_n) + \sgn(\omega_n{+}\Omega_m)\big]
    \int\frac{d\,\delta \tilde{k}_y}{2\pi}\;\frac{1}{W(i\omega_n) + W(i(\omega_n{+}\Omega_m)) - \Omega_0(\bq) - \kappa \delta \tilde{k}_y^2}\, .
\end{equation}
Taking $\Omega_m > 0$ allows us to write ($\tilde{\omega}_n \equiv -\omega_n - \Omega_m$)
\begin{equation}
    \begin{aligned}
        \sum_{\omega_n}\big[\sgn(\omega_n) + \sgn(\omega_n{+}\Omega_m)\big] f(\omega_n)
    &= \sum_{\omega_n > 0}\big[2 \big] f(\omega_n) + \sum_{\omega_n < -\Omega_m}\big[-2 \big] f(\omega_n)
    \\
    &= \sum_{\omega_n > 0}\big[2 \big]f(\omega_n) + \sum_{\tilde{\omega}_n > 0}\big[-2 \big]f(-\tilde{\omega}_n - \Omega_m)
    \\
    &= 2\sum_{\omega_n > 0}f(\omega_n) - f(-{\omega}_n - \Omega_m).
    \end{aligned}
\end{equation}
This allows us to write
\begin{equation}
    \begin{aligned}
        \Pi_{\rm{U}}(\bq, i\Omega_m) 
    &= \frac{ig^2 T}{v_F}
    \sum_{\omega_n > 0}
    \int\frac{d\,\delta \tilde{k}_y}{2\pi}\;\frac{1}{W(i\omega_n) + W(i(\omega_n{+}\Omega_m)) - \Omega_0(\bq) - \kappa \delta \tilde{k}_y^2} 
    - \frac{1}{ W(-i(\omega_n{+}\Omega_m)) + W(-i\omega_n) - \Omega_0(\bq) - \kappa \delta \tilde{k}_y^2}\,
    \\
    &= \frac{ig^2 T}{v_F}
    \sum_{\omega_n > 0}
    \int\frac{d\,\delta \tilde{k}_y}{2\pi}\; (2i) \text{Im} \left[ \frac{1}{W(i\omega_n) + W(i(\omega_n{+}\Omega_m)) - \Omega_0(\bq) - \kappa \delta \tilde{k}_y^2}  \right]\,
    \\
    &= -\frac{2g^2 T}{v_F} \text{Im} \left[ \sum_{\omega_n > 0}
    \int\frac{d\,\delta \tilde{k}_y}{2\pi}\;\frac{1}{W(i\omega_n) + W(i(\omega_n{+}\Omega_m)) - \Omega_0(\bq) - \kappa \delta \tilde{k}_y^2}  \right]
    \\
    &= -\frac{2g^2 T}{v_F} \text{Im} \left[ \sum_{\omega_n > 0}
     \frac{-i}{2 \sqrt{ \kappa (W(i\omega_n) + W(i(\omega_n{+}\Omega_m)) - \Omega_0(\bq))}} \right]
     \\
     &= \frac{g^2 T}{v_F \sqrt{\kappa}} \text{Re} \left[ \sum_{\omega_n > 0}
     \frac{1}{\sqrt{ W(i\omega_n) + W(i(\omega_n{+}\Omega_m)) - \Omega_0(\bq)}} \right]
    \end{aligned}
\end{equation}
where we've used the fact that $W(-i \omega_n) = W^*(i \omega_n)$ since $W$ is purely imaginary and in performing the contour integral over $\delta \tilde{k}_y$, we've used the fact that $\text{Im}[W(i\omega_n) + W(i(\omega_n{+}\Omega_m)) - \Omega_0(\bq)] > 0$ for $\omega_n, \Omega_m > 0$.
The above expression gives insight into the origin of the threshold discussed in Sec.\ref{subsec:umklapp}.
In the extreme case of large $\Omega_0(\bq),$ the piece in the summation is almost purely imaginary dominated by $-\Omega_0(\bq)$ inside the square-root for small $\omega_n$ so the real part becomes negligible and at large $\omega_n,$ the $\Omega_m$ dependence is washed away anyways.
The summation over $\omega_n$ is divergent and the physically meaningful $\Omega_m$ dependent piece is captured by $\Delta \Pi_{\rm{U}}(\bq, i\Omega_m) \equiv \Pi_{\rm{U}}(\bq, i\Omega_m) - \Pi_{\rm{U}}(\bq, 0)$.

To gain insight into the scaling behavior of $\Pi_{\rm{U}}$ we take $\delta q_x = q_y=0 \implies \Omega_0(\bq) = 0$, and approximate the Matsubara summation with continuous integrals to obtain

\begin{equation}
\label{eq:Pi_after_dkx}
    \Delta \Pi_{\rm{U}}(\bq=(\Delta_q, 0), i\Omega) 
    \propto \text{Re} \left[ \int_0^\infty d \omega \, [W(i\omega) + W(i(\omega {+} \Omega))]^{-1/2} - [2W(i\omega)]^{-1/2} \right],
\end{equation}

We first approximate $W(i\omega) \approx -\Sigma(i\omega) \sim i B\omega^{2/3}$ deep in the non-Fermi liquid regime.
Then we rescale $\omega = \Omega \tilde{\omega}$ to find immediately that
\begin{equation}
    \Pi_{\rm{U}}(\bq=(\Delta_q, 0), i\Omega) \sim  \frac{g^2\Omega^{2/3}}{v_F \sqrt{\kappa B}} ,
\end{equation}
which is what we discussed in Sec.[\ref{subsec:Bottlenecks}].
At higher temperatures it is safe to approximate $W(i\omega) \approx i \omega$ in which case the frequency rescaling remains the same $\omega = \Omega \tilde{\omega}$ giving
\begin{equation}
    \Pi_{\rm{U}}(\bq=(\Delta_q, 0), i\Omega) \propto \sqrt{\Omega}.
\end{equation}

We also discuss what happens in three dimensions.
When all is said and done, we end up with an expression that looks like
\begin{equation}
    \Pi_{\rm{U}}^{\rm{3D}}(\bq=(\Delta_q, 0), i\Omega_m) 
    \propto  \int d\omega \cdots\int\frac{d\,\delta k_y \; d\,\delta k_z}{2\pi}\;\frac{1}{W(i\omega_n) + W(i(\omega_n{+}\Omega_m)) - \kappa\,\delta k_y^2 - \kappa\,\delta k_z^2}
\end{equation}
which on moving to polar coordinates becomes
\begin{equation}
    \Pi_{\rm{U}}^{\rm{3D}}(\bq=(\Delta_q, 0), i\Omega_m) 
    \propto  \int d\omega \cdots\int \;\frac{r dr}{W(i\omega_n) + W(i(\omega_n{+}\Omega_m)) - r^2}
\end{equation}
which is now a simple logarithmic integral.
How $W(i \omega)$ scales with $\omega$ doesn't influence how $\Pi_{\rm{U}}^{\rm{3D}}(i\Omega)$ scales with $\Omega$.
We still need to rescale $\omega = \Omega \tilde{\omega}$ and that gives us back the same result as fond in Sec.[\ref{sec:3D}]
\begin{equation}
    \Pi_{\rm{U}}^{\rm{3D}}(\bq=(\Delta_q, 0), i\Omega) \propto \Omega.
\end{equation}

\section{Scattering Rate Estimate}
\label{app:B}

\begin{figure}
	\centering
		\includegraphics[scale=1.0]{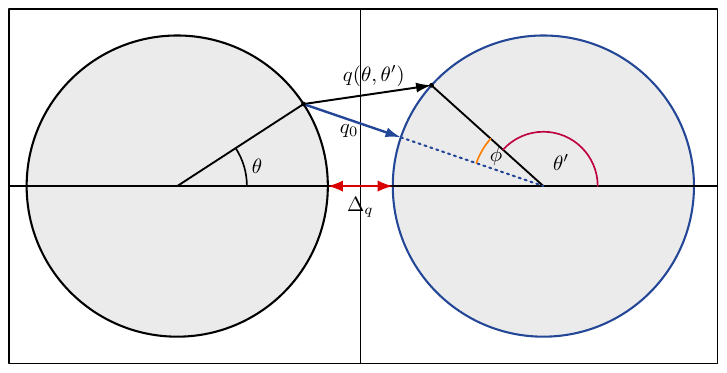}
	\caption{The parameterization of the FS used to evaluate the scattering rate.
    The point $\textbf{K}_F^x$ corresponds to the electron at $\theta=0$ which sets $\theta' + \phi = \pi$.
    For this electron the scattering momenta $q_\phi \equiv q(\theta=0, \theta' = \pi - \phi)$}
	\label{fig:Lee_Umklapp}
\end{figure}

We start with the dressed boson propagator
\begin{equation}
    D^{-1}(\bq, \Omega) = q^2 + aT - \eta\frac{i \Omega}{q} - i \lambda \frac{\Omega}{\sqrt{\Omega - \Omega_0(\bq)}} ,
\end{equation}
from which we can immediately write down the spectral function as
\begin{equation}
    B(\bq, \Omega) 
    = \frac{\eta\Omega/q + \lambda \frac{\Omega}{\sqrt{\Omega - \Omega_0(\bq)}}}{(q^2 + aT )^2 + (\eta\Omega/q + \lambda \frac{\Omega}{\sqrt{\Omega - \Omega_0(\bq)}})^2}.
\end{equation}

We reuse the scattering rate estimate formula from \cite{Lee_umklapp}
\begin{equation}
\label{eq:Scattering_Rate}
    \tilde{\gamma}(T) 
    \propto T \int_0^T d \Omega \int_{-\phi_{\rm{max}}}^{\phi_{\rm{max}}} d \phi \ \frac{B(q_\phi, \Omega)}{\Omega} 
    =  T \int_0^T d \Omega \int_{-\phi_{\rm{max}}}^{\phi_{\rm{max}}} d \phi  \frac{\eta \,q_{\phi}^{-1} + \lambda (\Omega - \Omega_0(\bq))^{-1/2}}{(q^2 + aT )^2 + \Omega^2(\eta \,q_{\phi}^{-1} + \lambda  (\Omega - \Omega_0(\bq))^{-1/2})^2}.
\end{equation}

We choose $T_3$ as the temperature scale relative to which we evaluate the dimensionless scattering in terms of $x \equiv T/T_3$ rate which gives us
\begin{equation}
\label{eq:Reduced_Scattering_Rate}
    \tilde{\gamma}(x) 
    = 
    x \int_0^x d \tilde{\Omega} \int_{-\phi_{\rm{max}}}^{\phi_{\rm{max}}} d \phi  
    \frac{\alpha^3 \,\tilde{q}_{\phi}^{-1} + \alpha^{3/2} (\tilde{\Omega} + \alpha^{-3} \sin^2(\phi)/4)^{-1/2}}{(\tilde{q}_{\phi}^{2} + \alpha_M x )^2 + \tilde{\Omega}^2(\alpha^3 \,\tilde{q}_{\phi}^{-1} + \alpha^{3/2} (\tilde{\Omega} + \alpha^{-3} \sin^2(\phi)/4)^{-1/2})^2}.
\end{equation}
In the above formula $\tilde{\Omega} \equiv T_3 \Omega, \; \tilde{q}_{\phi} \equiv q_{\phi}/k_F, \; q_\phi^2 \equiv q(\theta=0, \theta' = \pi - \phi) = k_F^2 + (\Delta_q + k_F)^2 - 2k_F(k_F + \Delta_q) \cos(\phi), \; \alpha \equiv [\Delta_q/k_F]/ [g^2 \varepsilon_F^{-1}]$ and $\cos(\phi_{\rm{max}}) \equiv k_F/(k_F + \Delta_q)$.

We also need to use the fact that $\Omega_{0}(\bq) \equiv \vF(\Delta_q - q_x) + \frac{\kappa}{4}q_y^2\, = \tilde{\varepsilon}_{\bq} - \frac{\kappa}{4}q_y^2$.
When the scattering out electron lies exactly on the umklapp FS, we have $\tilde{\varepsilon}_{\bq} = 0 \implies -\Omega_{0}(\bq) = \frac{\kappa}{4} k_F^2 \sin^2(\phi)$ since $q_y = k_F \sin(\phi)$.

The analogous formula with the boson thermal mass but without $\Pi_{\rm{U}}$ from \cite{Lee_umklapp} is
\begin{equation}
\label{eq:Reduced_Scattering_Rate_Lee}
    \tilde{\gamma}(x) 
     =
     x \int_0^x d \tilde{\Omega} \int_{-\phi_{\rm{max}}}^{\phi_{\rm{max}}} d \phi  
    \frac{\alpha^3 \,\tilde{q}_{\phi}^{-1}}{(\tilde{q}_{\phi}^{2} + \alpha_M x )^2 + \tilde{\Omega}^2(\alpha^3 \,\tilde{q}_{\phi}^{-1})^2}.
\end{equation}
The value of $\alpha_M$ is set by the specific model of our choosing.
For the SYK model, since we set $T_m = T_4$ this gives us $\alpha_M = \alpha$.
For the naive large-$N$ model, since we set $T_m = T_2$ this gives us $\alpha_M = \alpha^3$.

\end{widetext}

\renewcommand*{\bibfont}{\footnotesize}


\bibliography{main.bib}
\end{document}